\documentclass{aa}
\usepackage{txfonts}
\usepackage{graphicx}

\begin{document}

%\title{VINCI-VLTI measurements of HR 4049: the physical size of the circumbinary envelope
%	\thanks{Based on public shared risk science program data released from the ESO VLTI: 
%	http://www.eso.org/projects/vlti/instru/vinci/vinci\_data\_sets.html.}}	

\title{VINCI-VLTI measurements of HR 4049: the physical size of the circumbinary envelope
	\thanks{Based on public shared risk science program data released from the ESO VLTI: 
	http://www.eso.org/projects/vlti/instru/vinci/vinci\_data\_sets.html.}}

\author{S. Antoniucci 
        \inst{1,2,3},
	F. Paresce 
	\inst{3}
	\and 
	M. Wittkowski
	\inst{3}
	}

\institute{	Universit\`a degli Studi di Roma `Tor Vergata', via della Ricerca Scientifica 1, I-00133 Roma, Italy \and 
		INAF-Osservatorio Astronomico di Roma, Via di Frascati 33, I-00040 Monteporzio Catone, Italy \and
		ESO-European Southern Observatory, Karl-Schwarzschild-Stra\ss e 2, D-85748 Garching bei M\"unchen, Germany }	

%\offprints{Simone Antoniucci, present address:\\ santoniu@eso.org}
\offprints{Simone Antoniucci, email address:\\ antoniucci@mporzio.astro.it}

\date{Received date / Accepted date}

\abstract{
We present the first detection of the envelope which surrounds the post-AGB binary source HR 4049.
VINCI-VLTI $K$-band interferometric observations of this source imply the existence of a large structure with a Gaussian 
angular FWHM 22.4 $\pm$ 1.4 mas or uniform disk diameter of 34.9 $\pm$ 1.9 mas. With the Hipparcos 
parallax of 1.50  $\pm$ 0.64 mas these values correspond to a physical size of $14.9^{+11.1}_{-4.4}$ AU and $23.3^{+17.3}_{-7.0}$ 
AU, respectively. Our measurements, covering an azimuth range of $\sim$ 60\degr\, for the sky-projected baseline, provide information on 
the geometry of the emitting region and show that there is only a slight variation of the measured angular values along the 
different directions sampled. 
Thus, our results are consistent with a spherical geometry of the envelope. However, we cannot completely rule out the existence of 
an asymmetric envelope (like the circumbinary disk envisaged by some recent models) because of the limited spatial frequency and azimuth 
range covered by the observations.
	 }

\titlerunning{The physical size of HR 4049 envelope}
\authorrunning{Antoniucci et al.}

\maketitle

\section{Introduction}
HR 4049 (HD 89353, AG Ant) is considered the prototype of a peculiar group of sources in the post-Asymptotic Giant Branch 
(AGB) stage of evolution that are believed to be close binary systems surrounded by a massive and stable circumbinary envelope 
as a consequence of mass loss processes during the AGB phase.
These sources typically show evidence of extremely metal-depleted photospheres, possibly the result of dust formation in 
the circumstellar environment followed by re-accretion  of the depleted gas onto the stars.
In the case of HR 4049 the presence of large amounts of circumstellar dust is indicated both by the UV deficiency and 
strong IR excess (\cite{lam}). At the moment, however, the size and spatial distribution of this material are not well known.

Several models of HR 4049 have been proposed, mainly based on photometric and polarimetric observations, with different
predictions on the geometry and extent of the dust envelope, depending also on the physical properties of the dust grains.
Waters et al. (1989)\nocite{wat} suggested, for instance, that the observed IR excess could be modelled by the emission 
from an optically thin cloud with a radius ranging from 5 to about 300 AU or alternatively by a blackbody-emitting structure 
with a size of the order of 2 AU.
More recent work on HR 4049 tended to favour the flattened or asymmetric disk geometry for the distribution of the dust 
around the source.
Photometric variations in phase with variations of radial velocity indeed suggest that the circumstellar matter is 
distributed in a disk surrounding the binary, which would be observed from Earth at a high inclination with respect to the 
plane of the sky (\cite{wae}).
Analysis of the spectral energy distribution (SED) of HR 4049 has shown that the infrared excess between 1 and 850 $\mu$m 
can be fitted almost perfectly with a single blackbody function at $T \sim 1150\,$ K. On this basis, it has been suggested 
that the radiation we observe is coming from the inner rim of a circumbinary disk heated by the central sources (\cite{dom}). 
Moreover, polarization measurements indicate the presence of Rayleigh scattering by two different populations of grains, 
suggesting that the distribution of the circumstellar material might be asymmetric with the presence of a disk and a bipolar 
structure (\cite{jos}; \cite{jj}; \cite{joh}). 

In this paper, we present the analysis of VLTI-VINCI observations intended to shed light on this interesting problem by 
measuring the size and shape of the circumstellar envelope of HR 4049 directly by interferometric means. The observations clearly 
resolve for the first time a relatively large structure around the source with a dimension and shape that is quite plausibly 
attributable to the putative circumbinary envelope surrounding the binary.

\section{Observations and data reduction}
The observations (see Tab.\ref{nights} for a summary) have been carried out with the VLT Interferometer (e.g. \cite{gli}) 
between December 2002 and February 2003 using the near-infrared $K$-band VLTI commissioning instrument VINCI (e.g. \cite{ker1}), 
which recombines the light coming from two 0.35 m aperture siderostats (\cite{der}). 
The data were taken in the framework of the VLTI shared risk science program and are publicly available from the ESO 
Science Archive. The two interferometric baselines B3-C3 and B3-D1 were used, which have a ground length of 8 and 24 meters, 
respectively, and the same ground position angle (East of North) of 71\degr.

With this configuration, several series of interferograms (consisting of 100, 200 and 500 scans) were obtained on 
HR 4049, with a fringe frequency of 60 Hz. This rather slow frequency had to be chosen because of the relatively low $K$-magnitude 
of the object (K = 3.21 $\pm$ 0.29, \cite{cut}).
The details about the calibrator stars used for the observations are shown in Tab.\ref{calib}.
The fringe frequency for several calibrators was 290 Hz, different from the one used on the science target. This 
may introduce a calibration error which has been taken into account.
By comparing the results obtained using only calibrators sampled with the same fringe frequency as the target (whenever possible) 
to those derived using all available calibrators, we have estimated the effect of this error on the calibrated visibilities to 
be less than 3\%.

%***************************************************************************************************************************

\begin{table}[!t]

\centering

\caption[]{Observational log with nights, baselines and sky-projected baseline azimuth for VINCI observations of HR 4049.} 

\begin{tabular}{l l l l l}
\hline
Night & \multicolumn{2}{c}{VLTI Baseline} & Proj. Baseline (m) & Azimuth  \\                             
\hline
2002-12-16  &B3--C3 & 8 m & 7.4--7.5 &53$^{\circ}$--62$^{\circ}$\\
2003-01-13  &B3--C3 & 8 m & 7.8--7.9 &75$^{\circ}$--76$^{\circ}$\\
2003-01-15  &B3--C3 & 8 m & 7.8--8.0 &64$^{\circ}$--76$^{\circ}$\\
2003-01-16  &B3--C3 & 8 m & 7.9 &74$^{\circ}$--75$^{\circ}$\\
2003-01-20  &B3--C3 & 8 m & 7.5--7.8 &77$^{\circ}$--80$^{\circ}$\\
2003-02-25  &B3--D1 & 24 m & 11.3--12.3 &105$^{\circ}$--108$^{\circ}$\\
\hline                                                                          

\end{tabular}

\label{nights}
\end{table}

%***************************************************************************************************************************

Squared coherence factors were calculated processing the raw interferograms with the VINCI data reduction software 
(version 3.0) by Kervella et al. (2004).\nocite{ker2} 
Subsequent calibration of the data and calculation of broad-band synthetic visibility values were performed as in 
Wittkowski et al. (2004)\nocite{wit}.

A total of 17 values for the squared visibility were finally obtained from data reduction at an effective wavelength of 
$\lambda_\mathrm{eff}\:=\:2.21\:\mu$m and in an azimuth range for the sky-projected baseline between 53\degr\,and 
108\degr (North through East).

We use two simple two-component models for the object intensity distribution, both composed of a single central star 
(the primary of the binary system) sourrounded by an envelope. 
For the central star, we use the radius $R_*=47 R_{\sun}$ derived by Bakker et al. (1998) \nocite{bak} 
and adopt an almost unresolved uniform disk (UD) brightness distribution with angular 
diameter 0.7 mas (on the basis of the Hipparcos-determined distance).
Since the flux coming from the source is mainly due to the primary (e.g. \cite{bak}) we do not take into account 
the much fainter secondary which has been detected by radial velocity measurements (\cite{wae}; \cite{bak}).
As for the envelope, we describe it with an UD and a Gaussian distribution.
We assume a ratio $F = F_\mathrm{*} / F_\mathrm{tot} \sim$ 15\% for the stellar flux with respect to the 
total observed flux in the $K$-band (\cite{dom}).
Keeping the adopted parameters of the (unresolved) star constant and treating the UD diameter or the Gaussian FWHM 
of the envelope as the only free parameter, we compare the observed visibility with our simple model.

Although these simple two-component models certainly represent a first approximation, this procedure can provide 
us basic information on the shape and dimensions of the emitting region in HR 4049 without having to resort to more complex 
and uncertain assumptions. We should bear in mind that interferometer observations carried out at different azimuth values 
for the projected baseline allow us, in principle, to detect asymmetries of the intensity distribution.

%***************************************************************************************************************************

\begin{table}[!t]
\begin{center}

\caption[]{Calibrator stars. The uniform disk diameter $\theta_\mathrm{UD}$ and the corrisponding error $\sigma_{\theta}$ 
are reported together with spectral type, K magnitude and effective temperature.}

\begin{tabular}{l l l l l l}

\hline
Star            &  Sp.Type & K mag & $\theta_\mathrm{UD}$  & $\sigma_{\theta}$ & $T_{eff}$  \\
                &          &       & mas             	   & mas               &          K \\
\hline

 31 Ori$^a$              & K5 III     &  0.90 & 3.56 & 0.057 & 4046  \\%& Bord\'e $^a$ \\
 51 Ori$^a$              & K1 III     &  2.10 & 1.87 & 0.021 & 4508  \\%& Bord\'e $^a$ \\
 $\alpha$ Hya$^b$        & K3II-III   & -1.21 & 9.11 & 0.911 & 4230  \\%& Dyck $^b$    \\
 $\delta$ Lep$^a$        & K0 III     &  1.37 & 2.56 & 0.041 & 4656  \\%& Bord\'e $^a$ \\
 $\epsilon$ Lep$^a$      & K5 IIIv    & -0.20 & 5.91 & 0.064 & 4046  \\%& Bord\'e $^a$ \\
 $\iota$ Hya$^a$         & K2.5 III   & 0.91  & 3.41 & 0.048 & 4318  \\%& Bord\'e $^a$ \\
 $\nu_{2}$ CMa$^a$       & K1.5 III/IV& 1.62  & 2.38 & 0.026 & 4497  \\%& Bord\'e $^a$ \\
 $\mu$ Hya$^b$           & K4 III     & 0.40  & 4.55 & 0.455 & 4090  \\%& Dyck $^b$    \\
 v337 Car$^a$            & K2.5 II    & 0.03  & 5.09 & 0.058 & 4300  \\%& Bord\'e $^a$ \\

\hline 

\end{tabular}
\end{center}

\vspace{0.2 cm}

\small{$^a$ from Bord\'e et al., 2002\nocite{bor}, A catalogue of calibrator stars for LBSI.}\\*
\small{$^b$from calibration by Dyck et al., 1996\nocite{dyck} (Radii and Effective Temperatures for K and M 
Giants and Supergiants), using K magnitudes 
from Gezari, 1999\nocite{gez} (Catalog of IR observations).}\\*

\vspace{-0.5 cm}
\label{calib}
\end{table}

%***************************************************************************************************************************

\begin{figure*}[!t]
\includegraphics[width=6cm]{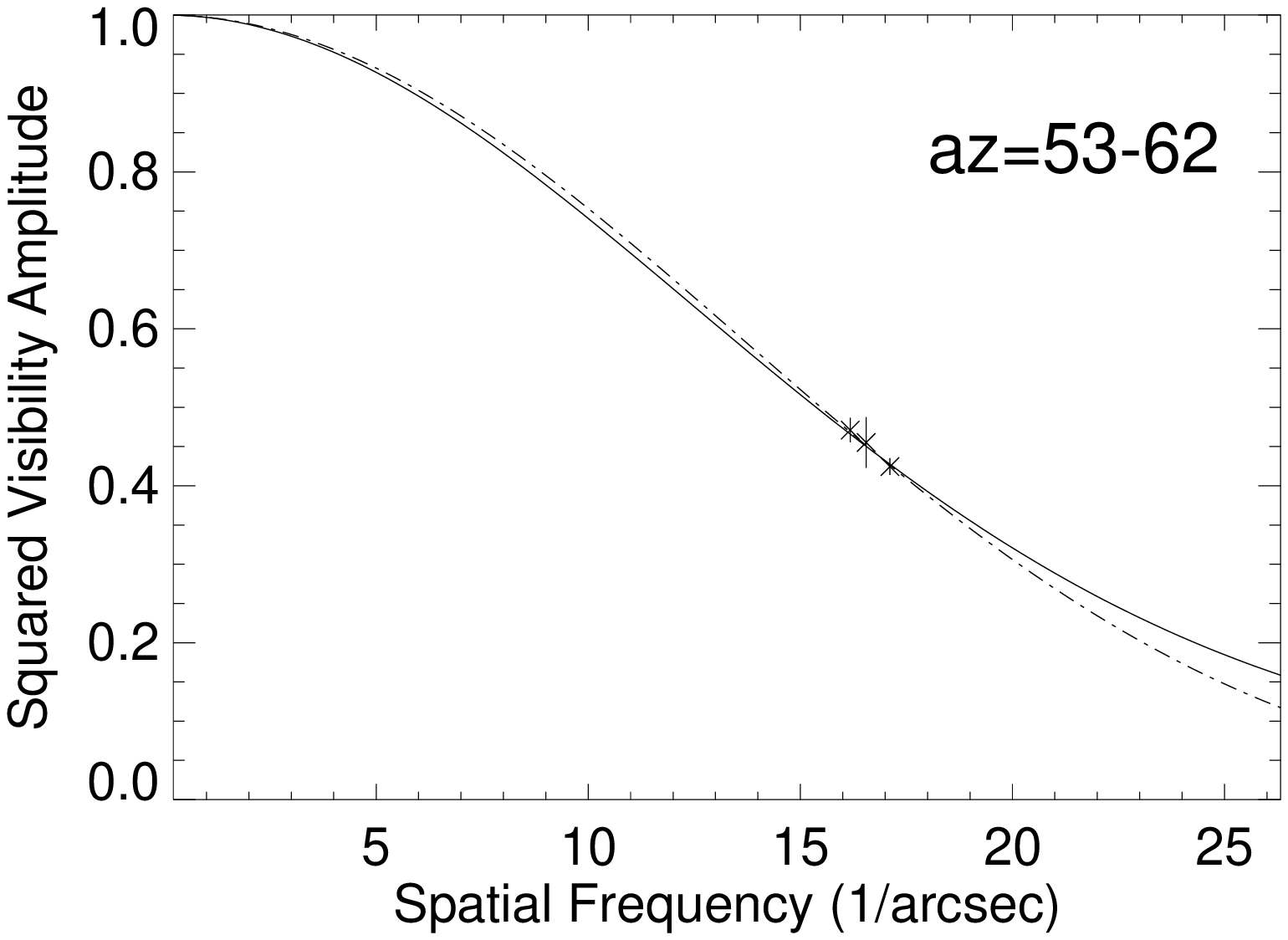}%
\includegraphics[width=6cm]{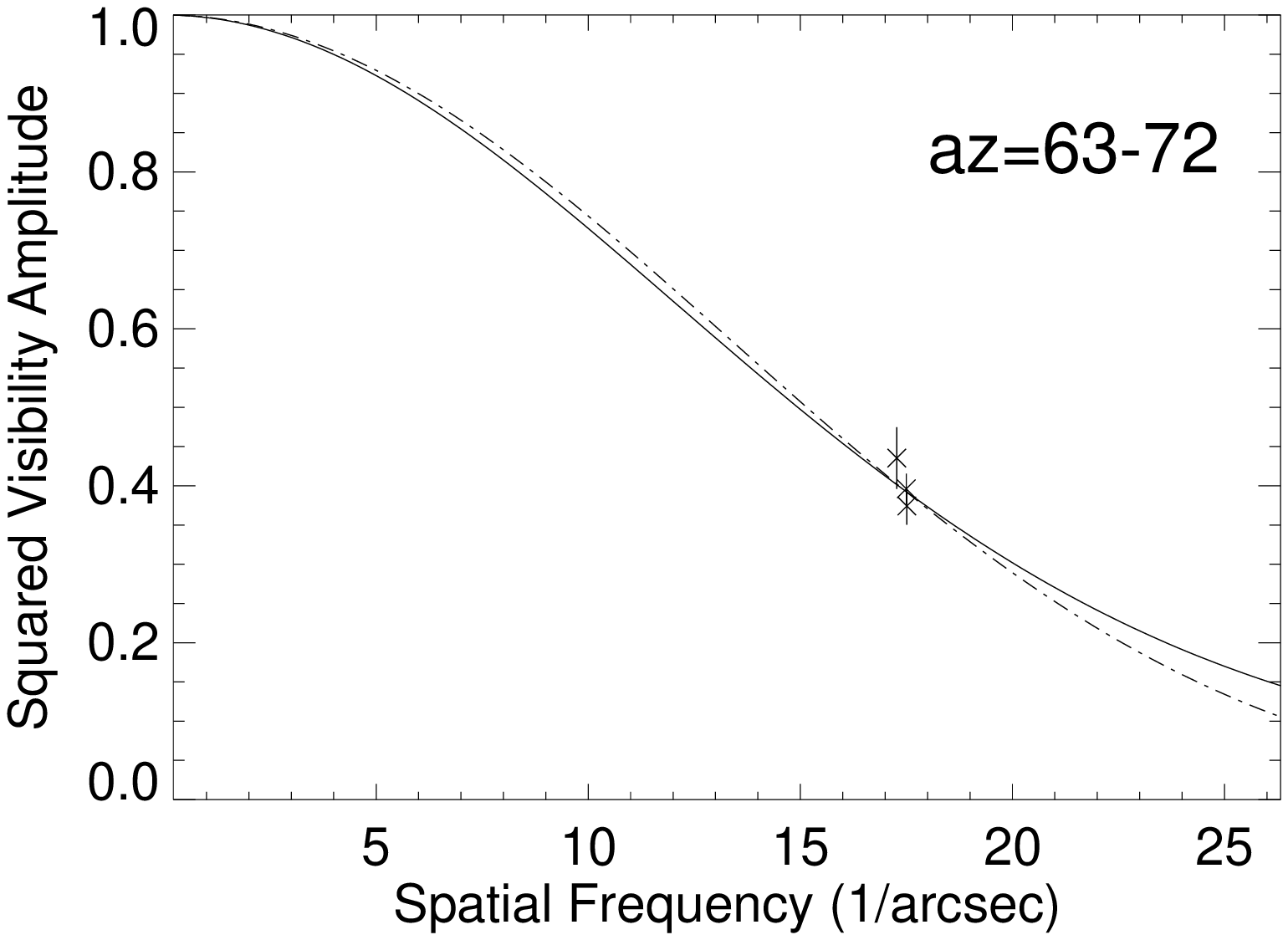}\\
\vspace{-0.5cm}
%\sidecaption
\includegraphics[width=6cm]{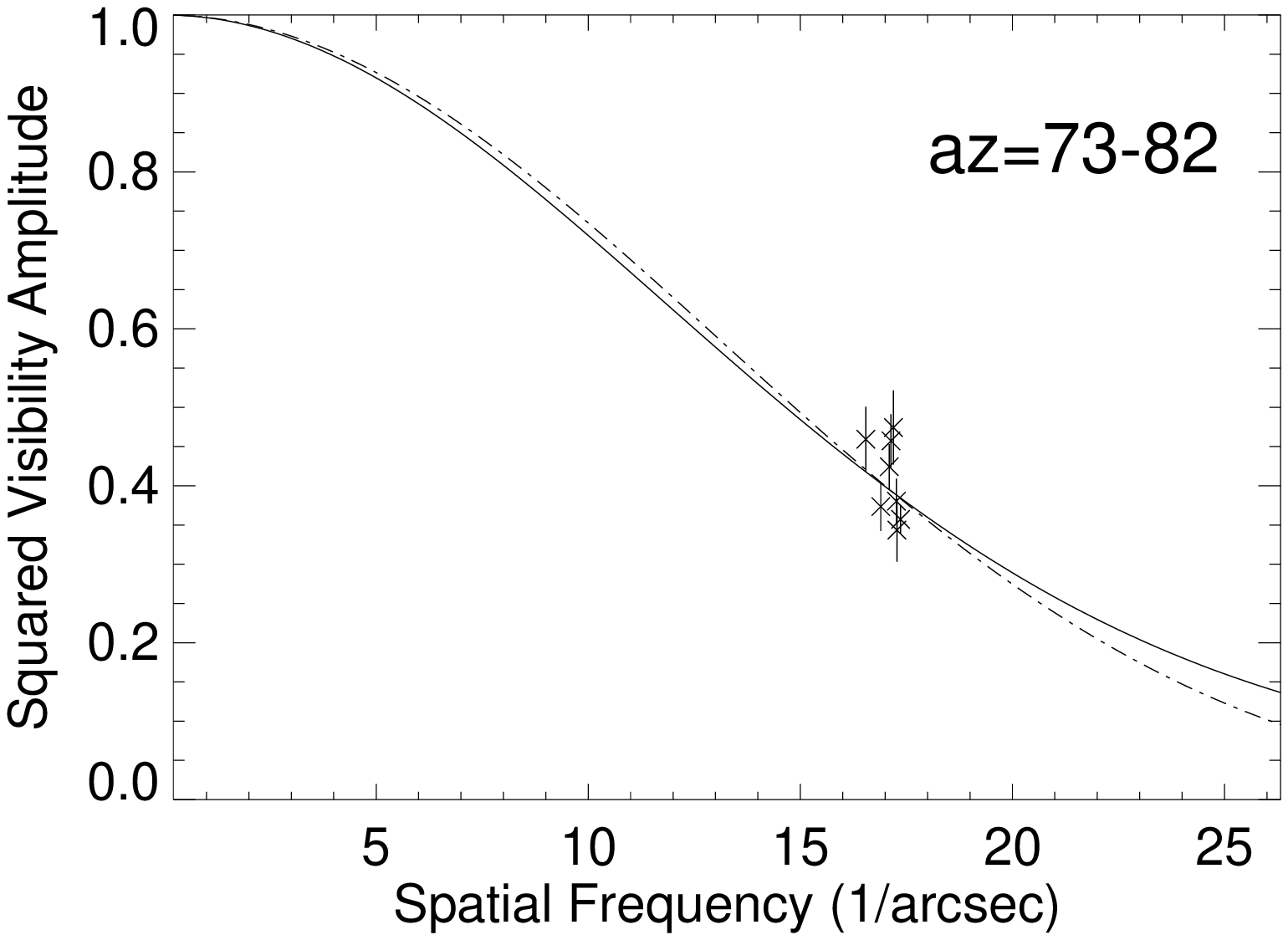}%
\includegraphics[width=6cm]{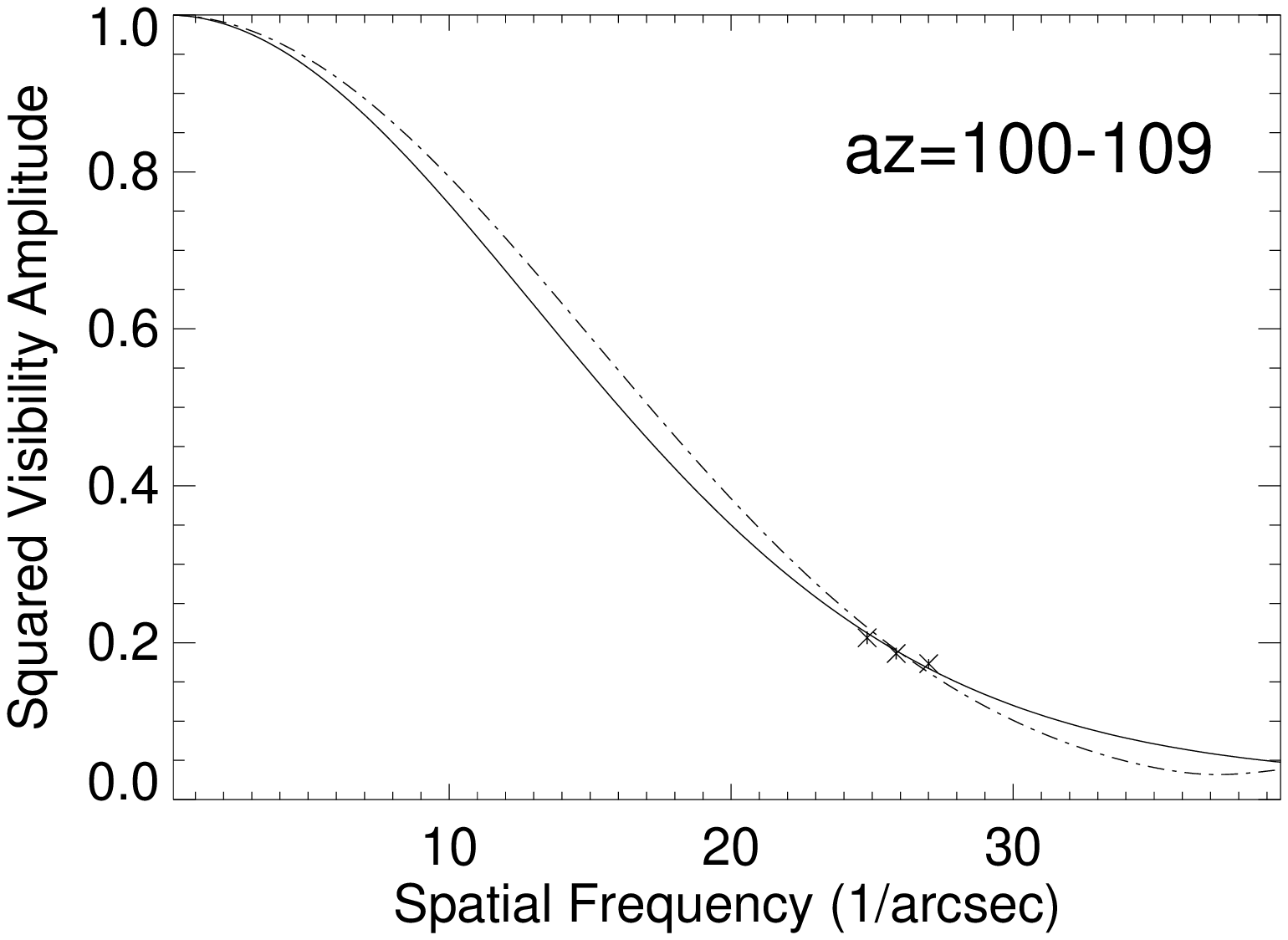}
\vspace{0.5cm}
\caption{Best-fit through the observed visibility points of the normalized squared visibility curve provided by 
the simple models of HR 4049 described 
in the paper. The solid curve corresponds to the UD + Gaussian (star + envelope) model, the dashed-dotted line to 
the UD + UD model, both for a stellar to total flux 
ratio $F$=15\%. The four panels refer to the different bins of azimuth (see Table~\ref{newbins}).} 
\label{plots} 
\end{figure*}

For this reason, we grouped the data points in four different 10\degr\, wide azimuth bins (53\degr--62\degr, 63\degr--72\degr, 
73\degr--82\degr\, and 103\degr--112\degr). 
Then, we performed a best-fit of our models to the squared visibility values for each bin to obtain the corresponding values 
of the Gaussian FWHM and UD diameter of the envelope (see Fig.~\ref{plots} and Tab.\ref{newbins}). 

We also investigated how the obtained results vary depending on the adopted flux ratio $F$. We found that,
when changing $F$ by 5\%, the related variation for the envelope angular size is small: about 5\% and 4\% for 
the Gaussian and UD distribution, respectively. 

Finally, although the decrease of the observed visibility between spatial frequencies $\sim$ 16 and $\sim$ 26 
cycles/arcsec clearly shows that 
we are resolving a compact envelope component, a contribution to the total visibility by an additional extended scattering component 
is quite possible. In this case, assuming that 10\% of the total flux is due to this incoherent component, we found that  
the fitted sizes of the envelope (UD and Gaussian) decrease by roughly 10\%.

%***************************************************************************************************************************

\begin{table}[!t]
\centering

\caption[]{VLTI-VINCI measurements of the size of the envelope surrounding HR 4049. Depending on the adopted 
source model (UD + Gauss or UD + UD, see the text), the FWHM of a Gaussian distribution or an equivalent uniform 
disk diameter $\theta\,^\mathrm{env}_\mathrm{UD}$ are reported. Formal errors are shown;
further error ($\sim$3\%) is introduced by the calibration process (see the text). The measurements refer to the 
four azimuth bins mentioned in the paper and to a value $F$=15\% of stellar to total flux ratio. The last row 
shows the results when considering all data together.}

\begin{tabular}{l | l | l}

\hline
Azimuth			& UD + Gauss 	 & UD+UD \\
\hline
Bins 	 		&  FWHM $\pm \sigma$ (mas)&  $\theta\,^\mathrm{env}_\mathrm{UD} \pm \sigma $ (mas)\\
\hline
53\degr--62\degr	& 22.7 $\pm$ 0.1 & 36.9 $\pm$ 0.1 \\
63\degr--72\degr	& 23.3 $\pm$ 0.4 & 37.7 $\pm$ 0.6 \\
73\degr--82\degr	& 23.8 $\pm$ 0.6 & 38.4 $\pm$ 0.8 \\ 
103\degr--112\degr	& 21.7 $\pm$ 0.2 & 33.4 $\pm$ 0.4 \\
\hline
all azimuths		& 22.4 $\pm$ 0.5 & 34.9 $\pm$ 0.7 \\
\hline
\end{tabular}
\label{newbins}
\vspace{-0.2cm}
\end{table}

%**************************************************************************

The measurements show that we have detected an emitting structure whose azimuthally averaged angular size is 22.4 
$\pm$ 1.4 mas (Gaussian FWHM) or 34.9 $\pm$ 1.9 mas (UD), with errors including also the previously mentioned uncertainties 
due to calibration and to the adopted flux ratio. 
There is only a slight variation of the size depending on baseline orientation, most of it when passing from the 
first three adjacent bins to the fourth one; the values range from 21.7 to 23.8 mas for the Gaussian and from 33.4 to 
38.4 mas for the UD distribution. 
This indicates that our observations are consistent with a roughly spherical symmetric envelope; however,  
having measurements for only two different baselines, and therefore in a limited range of spatial frequencies
and position angles, does not allow us to definitely rule out at this point the possibility that the envelope might be partly 
asymmetric and hence to unambiguously discriminate between a disk structure or a more spherical geometry.
A more flattened envelope could explain the small variability of the object in the $K$-band (about 0.1 mag in time scales of the 
order of 250 days, \cite{wae}); this would be due to different extiction along varying lines of sight to the primary 
as the components orbit each other (\cite{wae}).
Because of our limited data sample we can only give here an estimate of the envelope size and we cannot exclude the existence 
of more complex structures consisting of several clouds or regions with different optical depth.

With the Hipparcos parallax of 1.50 $\pm$ 0.64 mas (\cite{per}) the azimuthally averaged angular values obtained for 
the Gaussian FWHM and the UD diameter of the envelope (Tab.\ref{newbins}) correspond to linear dimensions 
of $14.9^{+11.1}_{-4.4}$ AU and $23.3^{+17.3}_{-7.0}$ AU, respectively.

\section{Discussion}

The geometry and extension shown by our results are in agreement with those proposed by
Waters et al. (1987)\nocite{wat}, who modelled the emission from HR 4049 as arising from a simple spherical 
optically thin envelope and found
that such structure would have a radius ranging from 5 AU up to about 300 AU.

Joshi et al. (1987)\nocite{jos} suggested that the observed intrinsic polarization of light from HR 4049 is introduced 
by scattering due to the circumstellar matter and that the low degree of polarization measured probably indicates that the 
envelope is almost spherically symmetric, which would be in agreement with our results. 
They modelled the strong dependence of the position angle 
with wavelength assuming the presence of two spheroidal dust shells surrounding the source, having different 
symmetry axis and composed by two different populations of large ($ a \sim 1 \mu$m) and smaller grains ($a \leq 0.05 \mu$m).

Johnson and Jones (1991)\nocite{jj} modelled the polarization as arising from scattering by dust in an envelope 
having the shape of a prolate ellipsoid; they found a position angle of about 8 degrees, 
remarking how this angle is approximately perpendicular to the direction of an extended (about 3 pc) emission at 
100 $\mu$m reported by Waters et al. (1989)\nocite{wat}, presumably due to cold matter ejected from the source 
during the previous evolutionary stage.

It is also of interest to compare our interferometric results with the model proposed by Dominik et al. (2003)\nocite{dom} 
in which the emission we observe would actually come from the inner surface of a geometrically ($ Height/Radius \sim$ 1/3) 
and optically thick circumbinary disk surrounding the central sources and seen at an inclination of $\sim$ 60\degr\, with 
respect to the plane of the sky. Using this model, the authors infer a radius of order $R \sim $ 10 AU 
for the distance of the inner surface of the disk from the star, which is roughly consistent with the size of the structure 
detected by VINCI. 
Because of the high inclination, however, we would expect to observe a somewhat larger variation in the measured 
size of the structure for different baseline orientations. Assuming that we are observing in an azimuth range 
around the major axis of the projected disk (this would reproduce the trend we observe in our measurements, with the smallest 
values for the first and last bin), we should in fact see a diameter variation of at least 4 mas, 
which is around the 3$\sigma$ level in our measurements.

Finally, Johnson et al. (1999)\nocite{joh} proposed that the measured optical polarization indicates Rayleigh 
scattering arising in an optically and geometrically thin circumbinary disk. This, together with varying extinction with 
orbital phase, has led the authors to formulate the hypothesis that the orientation of the disk is tied to the orbit of 
the binary and so the projected face of the disc would change as the central sources orbit each other.
Moreover, the $UV$ polarization shows a typical $90^{\circ}$ change in position angle, usually related to the presence 
of a bipolar structure (i.e. two structures orientated perpendicularly to each other), which is often seen in planetary nebulae.
In their measurements, the polarized light shows two preferential position angles (averaged over wavelength) 
of $\sim 17^\circ$ and $\sim 28^\circ$.
It is interesting to notice that, since the observed PA of scattered light is supposed to be perpendicular to 
the plane of the circumstellar material in the disk, the orientation angle of the long axis of the disk, as viewed from Earth, 
would lie in the range 110$^{\circ}$--120$^{\circ}$. Thus, the VINCI observations presented here (with $\alpha 
\sim$ 50$^{\circ}$--110$^{\circ}$) would sample directions ranging from approximately that of the (projected) disk's long axis 
($\alpha \sim$ 110$^{\circ}$) to one almost perpendicular.
However, our results show that there is no clearly evident difference in the source dimensions when passing from 
one azimuth bin to the others, as one might expect assuming the disk configuration reported above.

Although HR 4049 is generally believed now to be surrounded by a circumbinary disk on the basis of the different 
observational evidences mentioned above, the existence of a more spherical envelope should not be excluded out of hand, 
as the recent example of the Herbig AeBe stars clearly shows (e.g. \cite{mg}).

Moreover, it is also likely that the envelope around this source might be multi-structured, as the polarization 
data suggest (bipolar structure, different populations of grains in separated shells).

\section{Conclusions}
The reduction and analysis of VINCI data on HR 4049 have led to the detection of a structure of angular dimensions 
34.9 mas (uniform disk diameter) or 22.4 mas (FWHM of a Gaussian distribution), probably the circumstellar envelope 
or disk surrounding this binary system. 

Assuming the Hipparcos-determined distance of 667 pc for the source, this angular values corresponds to a size 
of about 23.3 AU and 14.9 AU respectively.

The measured size shows only a slight variation along different directions spanning an azimuth range of $\sim60^{\circ}$,
consistent with models of spherically symmetric or ellipsoidal envelopes 
(e.g. \cite{wat}; \cite{jos}; \cite{jj}).

We cannot, however, completely rule out the possibility of the presence of asymmetric 
structures, such as the circumbinary disk predicted by some recent models.

Our knowledge of the geometry of HR 4049 system can be refined with further interferometric observations. 
Sampling different azimuths and spatial frequencies and also carrying out measurements at different wavelengths would 
definitely help in constraining the models, thus providing new information on both the geometry and physics of 
the HR 4049 circumstellar envelope detected by VINCI. This will indeed be possible with the advent of the MIDI and AMBER 
instruments and the auxiliary telescopes to the VLTI facility in the next year or so.

\begin{acknowledgements}

We would like to thank I. Percheron, O.L. Lupie, V. Roccatagliata and D. Fedele for their helpful advice.
For this work we have made use of NASA's Astrophysics Data System Bibliographic
Services and of the SIMBAD database, operated at CDS, Strasbourg, France.

\end{acknowledgements}

%***************************************************************************************************************************

%***************************************************************************************************************************

\end{document}